\documentclass[preprint,12pt]{elsarticle}

\usepackage{listings}
\usepackage{amssymb}
\usepackage[T1]{fontenc}
\usepackage{algorithmic}
\usepackage{algorithm} 
\usepackage{array}
\usepackage{amsmath}
\usepackage{amsfonts}
\usepackage{amssymb}
\usepackage{amsthm}
\usepackage{caption}
\usepackage{comment}
\usepackage{fancyhdr}
\usepackage{geometry} 
\usepackage{graphicx}
\usepackage[colorlinks]{hyperref}
\usepackage{multicol}
\usepackage{multirow} 
\usepackage{rotating}
\usepackage{setspace}
\usepackage{graphicx}
\usepackage{url}
\usepackage{verbatim}
\usepackage{xcolor}
\usepackage{url}

\begin{document}

\begin{frontmatter}

\title{Unveiling Hybrid Cyclomatic Complexity: A Comprehensive Analysis and Evaluation as an Integral Feature in Automatic Defect Prediction Models}

\author[inst1]{Laura Diana Cernău}
\author[inst1]{Laura Dioșan}
\author[inst1]{Camelia Șerban}

\address[inst1]{Babeș-Bolyai University Faculty of Mathematics and Computer Science,\\ Kogălniceanu, 1, 400084, Cluj, Romania}

\begin{abstract}
The complex software systems developed nowadays require assessing their quality and proneness to errors. Reducing code complexity is a never-ending problem, especially in today's fast pace of software systems development. Therefore, the industry needs to find a method to determine the qualities of a software system, the degree of difficulty in developing new functionalities, or the system's proneness to errors. One way of measuring and predicting the quality attributes of a software system is to analyse the software metrics values for it and the relationships between them. More precisely, we should study the metrics that measure and determine the degree of complexity of the code. This paper aims to analyse a novel complexity metric, Hybrid Cyclomatic Complexity (HCC) and its efficiency as a feature in a defect prediction model. The main idea behind this new metric is that inherited complexity should play a role in the complexity of a class, hence the need for a metric that calculates the total complexity of a class, taking into account the complexities of its descendants.
Moreover, we will present a comparative study between the HCC metric and its two components, the inherited complexity and the actual complexity of a class in the object-oriented context. Since we want this metric to be as valuable as possible, the experiments will use data from open-source projects. One of the conclusions that can be drawn from these experiments is that inherited complexity is not correlated with class complexity. Therefore, HCC can be considered a valid metric from this point of view. Moreover, the evaluation of the efficiency of the prediction models shows us a similar efficiency for HCC and the inherited complexity. Additionally, there is a need for a clear distinction between a class's complexity and its inherited complexity when defining complexity metrics.
\end{abstract}


\end{frontmatter}


\section{Introduction}
\label{chapter:introduction}

Software complexity represents a subject that plays an essential role in most of the stages of the lifecycle of a project, such as the development, the testing or the maintenance phases. In addition to other uses, software complexity can predict software quality attributes like efficiency, correctness, maintainability, interoperability, reliability, portability, reusability or testability \cite{Misra2018}. 

Since the complexity of a software project can have implications in many areas of the development lifecycle, it is essential to know how to measure it so that the proper actions are taken in the early stages of a project. By doing this, we ensure that we minimise the risks of defects appearing in the final phases of the projects, involving unexpected costs and delays. An example that supports the previous statement is represented by the results of the systematic study presented in \cite{Nguyen2017}. The authors studied the impact of software complexity on software quality attributes. They concluded that software complexity metrics are some of the most used predictors for quality attributes such as fault proneness and maintainability. When dealing with a high degree of code complexity during the maintenance phase, the risk of introducing new defects increases, leading to higher costs and more extended periods of development. Furthermore, code complexity also influences code readability, with increased complexity leading to decreased readability. As code complexity increases, it becomes more challenging to extend the codebase effectively.

Therefore, a simple way to define software complexity would be how difficult it is to understand and modify a piece of code without introducing defects or unexpected behaviours. In his article, E.E. Ogheneovo gathers other definitions of software complexity, including several aspects of software complexity. Remarkably, he states that the more interdependent variables a system has, the more complex it is. Furthermore, he summarises the meaning of software complexity as the difficulty in understanding and testing the design and implementation of a system component \cite{Ogheneovo2014}.

Consequently, since the early detection of defects in a software system can significantly impact the final stages of the development process, we are analysing the impact of the inherited software complexity in the context of object-oriented programming in this paper. The Hybrid Cyclomatic Complexity metric (HCC) was proposed in \cite{Cernau2022}, and it defines the complexity of a class as the sum of all the inherited complexities from the class's parents. In introducing this metric, we intended to express that when assessing the complexity of a class, it should consider the complexity of its parent classes. This is particularly relevant in the context of object-oriented inheritance, as the child class inherits and utilises all the public and protected methods of its parent classes.
Furthermore, in this paper, we are further analysing the HCC and its effectiveness as a defect prediction compared to having the inherited complexity and the current class complexity as two different features for a prediction model. Moreover, one aspect worth mentioning is that there are many definitions for computing the complexity of a class, and not all of them mention if the inherited complexity is considered. We aim to make a clear distinction between the complexity of a class and its inherited complexity.

Our preliminary research suggests that there is not a strong relationship between the complexity of a class and the complexities inherited from its parents \cite{Cernau2022}. Therefore, this indicates that the validity of the HCC metric's definition originates from the fact that its two components are linearly independent. However, when studying the efficacy of HCC compared to its two parts taken separately as defect predictors, we cannot say that one approach stands out clearly compared to the other. Nevertheless, this paper shows that there is a probability that by summing the complexity of a class with the sum of the complexities of its parents under a single metric, important aspects regarding complexity are lost. 

This paper aims to study whether the HCC metric is comprehensive and complex, which does not minimise aspects of inheritance by summing up the inherited complexity with that of the child class. This metric will be used in the context of automatic defect prediction. Therefore, we want to validate its potential as a feature in a prediction model. Moreover, we want to emphasise that a clear distinction needs to be made between the complexity of a class and its inherited complexity when computing complexity-related software metrics.
To demonstrate the previously mentioned aspects, we will answer the following research questions:

\textbf{RQ1} How do the two features (the class's inherited and actual complexity) impact the automatic identification of software defects?

\textbf{RQ2} What inheritance-related aspects does the Hybrid Cyclomatic Complexity measure try to emphasise?


The HCC metric was defined to cover a broader spectrum of complexity in the context of automated software defect prediction. In the initial experiments \cite{Cernau2022}, the HCC metric performed well as a feature in a defect prediction model.

One of this paper's contributions lies in further analysing the effectiveness of the HCC metric and investigating its decomposition into the sum of inherited complexity and the actual complexity of the class. Additionally, this paper explores whether certain complexity aspects are minimised by adding the complexities of the parent classes to the actual class complexity. 
Lastly, we consider the experiments performed on three datasets a solid contribution to this research endeavour.

The remainder of this paper is structured as follows. First, the section \textit{Related work} (\ref{section:related-work}) represents a summary of the related work regarding the correlations between software quality attributes and software metrics. Next, section \textit{Problem definition} (\ref{section:problemDefinition}) describes in detail the problem that we are trying to solve and the main topics discussed in this paper. In section \textit{Investigated approach} (\ref{section:proposedApproach}), we illustrate the methodology used, the dataset description and exploratory data analysis. Moreover, the \textit{Experiments} section (\ref{section:experiments}) illustrates the experiments done for this paper and their breakdown. Furthermore, the \textit{Threats to validity} section (\ref{section:threats-to-validity}) discusses the possible threats to external and internal validity. Finally, the last two sections are reserved for discussions about future work (\ref{section:future_work}) and the conclusions of this paper (\ref{section:concl}).

\section{Related work}
\label{section:related-work}

Software quality attributes illustrate a software system's characteristics, usability, and performance. On the other hand, software metrics provide quantitative measurements that can offer insights about the software attributes displayed by a system. For instance, software metrics find application in quantifying software attributes (such as how code complexity affects maintainability) and predicting behaviours (defect prediction) during the refactoring process or when conducting root cause analysis. 

Consequently, multiple research papers examine the relationships between software complexity and quality attributes. For instance, in his article, Edward E. Ogheneovo studies software complexity's impact on maintenance costs. He begins from the premise that when the number of lines (LOC) increases, the complexity also increases. Therefore, after studying data related to operating systems (Windows, Debian Linux and Linux Kernel), he concluded that software complexity directly impacts maintenance costs. Therefore, efforts in measuring and controlling the software complexity of a system are justified \cite{Ogheneovo2014}.

Another example would be the correlation between testing effort (the number of test cases, the test execution time) and software complexity metrics. After their investigation, Muslija et al. concluded that software metrics moderately impact the testing effort. Furthermore, the size of the system proved to have the highest correlation with the testing effort \cite{Muslija2018}.

When Alfadel et al. analysed the measurement of the defect density using Halstead's metrics and the Cyclomatic Complexity metric, they concluded that both display equivalent associations to defect density. Moreover, due to their strong linear correlation, they can be used in measuring defect density \cite{Alfadel2017}.

Misra et al. propose in their paper \cite{Misra2018} a suite of object-oriented cognitive complexity metrics. The article compares the proposed metrics suite and the Chidamber and Kemerer metrics suite \cite{CK1994}. Moreover, they theoretically validate their suite using Weyuker's properties and empirically using Kaner's framework. Some of the conclusions of this article state that the proposed metric suite can be used to measure inheritance and coupling and to evaluate design efficiency in a software project \cite{Misra2018}.

In \cite{Masmali2021}, novel code complexity software metrics are introduced. The authors propose a set of complexity metrics that they claim to be more accurate in capturing software complexity than existing ones. In defining these new metrics, they consider aspects of coupling, cohesion, abstraction, and inheritance. Their motivation is based on the fact that traditional complexity metrics may fall short in assessing the complexity of modern systems, given the prevalent use of patterns and high-level abstractions. In their approach, they assign complexity values to each of the components of a class (attributes, methods, and associations) to calculate its overall complexity.

Numerous papers in the literature focus on the automated prediction of software defects. Below, we briefly describe some of these papers that share similarities with our approach.
In their article \cite{Fer2008}, the authors employ machine learning techniques to develop a predictor for classifying files as either clean or containing bugs. They utilise the decision tree algorithm and train the classifier with a dataset consisting of static code metrics and bug information associated with each file. The software metrics utilised in this study include NEL (Number of Executable Lines), CD (Control Density), CC (Cyclomatic Complexity), PC (Parameter Count), RP (Return Points), LVC (Local Variable Count), and ND (Nesting Depth).

Another relevant article is \cite{Alshehri2018}, where the authors assess the performance of a reduced set of static metrics, change metrics and a combination of both as predictors for identifying fault-prone code. Static code metrics include LOC (Lines of Code), Max Complexity, and Methods per Class. Change metrics encompass Ave-LOC Added, LOC-Deleted, Refactorings (indicating how many times a file was refactored), and code churn. The analysis uses three machine learning algorithms: Logistic Regression, Naive Bayes, and Decision Tree J48.

In \cite{Clemente2018}, a comparison is drawn between multilayer deep feedforward networks and traditional machine learning algorithms (decision tree, random forest, naive Bayes, and support vector machines) for predicting security-related faults using software quality metrics. The deep learning algorithm outperforms traditional methods in predicting security bugs. The metrics used in this research fall into three categories: object-oriented, complexity, and volume, with examples like Cyclomatic Complexity, Executable Statements, Declarative Statements, and Comment to Code Ratio.

A fault prediction model proposed in \cite{Kumar2017} combines three different ensemble learning algorithms. The authors found that a subset of code metrics significantly impacts prediction accuracy more than the entire code metrics set. This subset includes DIT (depth of inheritance tree), WMC(Weighted Methods per Count), CBO (Coupling between objects), LCOM3 (Lack of Cohesion of Methods), AVG-CC (Average Cyclomatic Complexity), and NOC (Number of Children), among other Chidamber and Kemerer Java Metrics.

\section{Problem definition}
\label{section:problemDefinition}

\begin{figure}[ht]
    \includegraphics[width=\columnwidth]{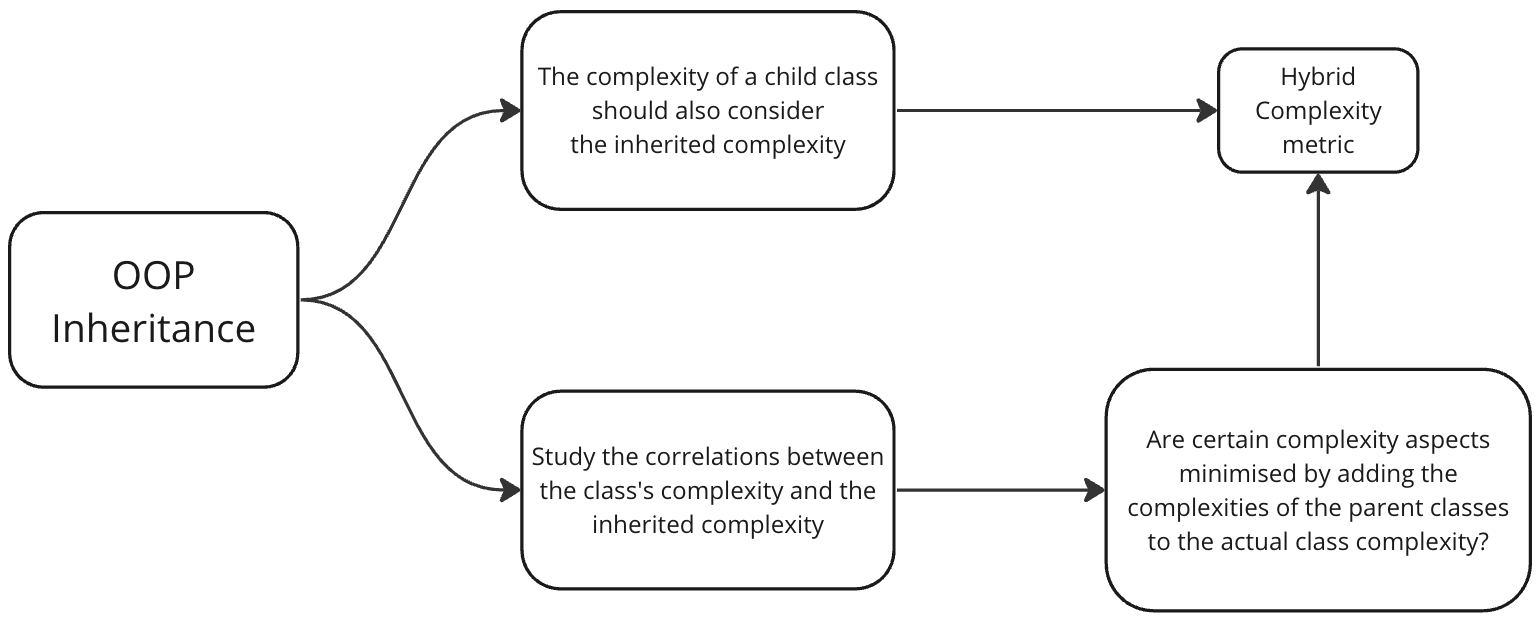}
    \caption[width=\columnwidth]{Hybrid Cyclomatic Complexity concepts and question map}
    \label{fig:problem-description}
\end{figure}

\begin{figure}[ht]
    \includegraphics[width=0.5\textwidth]{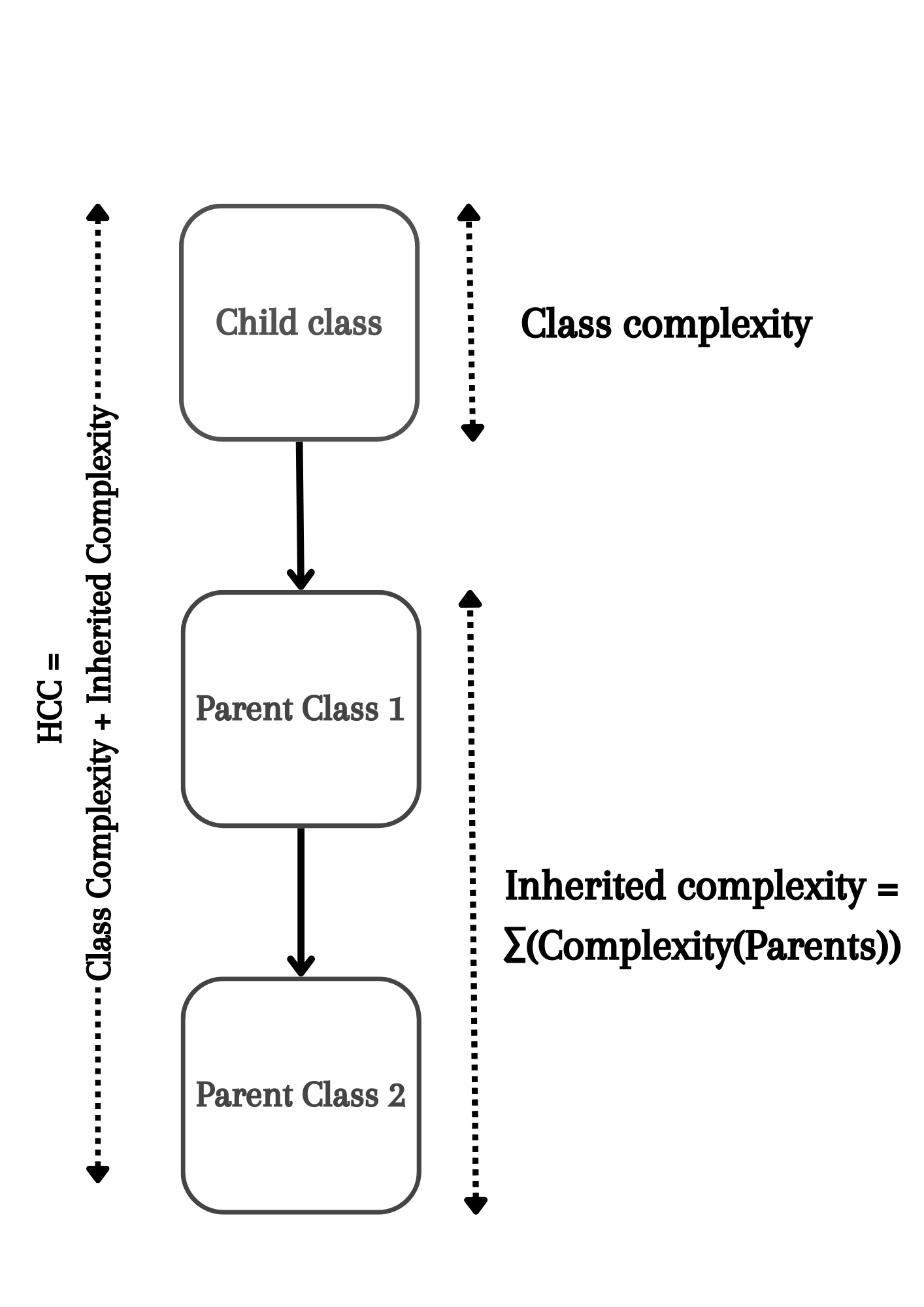}
    \centering
    \captionsetup{justification=centering}
    \caption{Hybrid Cyclomatic Complexity components}
    \label{fig:hcc-definition}
\end{figure}

Figure \ref{fig:problem-description} illustrates the essential considerations that led to the HCC metric definition and associated concerns. Therefore, inheritance is one of the most well-known object-oriented concepts. The initial intent of inheritance was to increase code reusability, thereby eliminating code duplication and facilitating reusability and extensibility. However, in the context of inheritance, the child class has access to all the protected attributes and methods of the parent class. Consequently, we assumed that the complexity of a child class should also encompass the inherited complexity. As a result, we introduced a complexity metric, HCC, which defines the complexity of a class as the sum of its complexity and all the complexities of its parents. In Figure \ref{fig:hcc-definition}, you can see the visual representation that illustrates the different components of the HCC metric.

In the initial experiments, the hybrid metric was a better-performing feature in defect prediction than the Weighted Method per Class (WMC)\cite{Kumar_2013}, which defines the complexity of a class. However, as we will demonstrate later in our paper, there is no clear correlation between the inherited complexity and the actual complexity of the class. Hence, we studied their performance as different features for a prediction model, and the efficiency was similar to the HCC's. Additionally, another concern is related to the definitions of the complexity of a class. More precisely, we do not know if the existing definitions consider inherited complexity.

\section{Investigated approach}
\label{section:proposedApproach}

\subsection{Glossary of Software Metrics}
\label{subsection:metrics-glossary}

For a better understanding of the proposed approach, this section starts with a review of the most important metrics used to determine the complexity of software projects.
 
\begin{table}[H]
  \centering
  \renewcommand{\arraystretch}{1.5}
  \begin{tabular}{|p{0.35\textwidth}|p{0.13\textwidth}|p{0.6\textwidth}|}
    \hline
    \textbf{Name} & \textbf{Acronym} & \textbf{Definition} \\
    \hline
    Cyclomatic Complexity & CC & The Cyclomatic Complexity metric quantifies the complexity of the code flow of a module, drawing its foundation from graph theory \cite{CK1994}. \\
    \hline
    Weighted Methods Per Count & WMC & The Weighted Methods Per Count quantifies the overall complexity of a class, obtained from summing all the complexities of its methods \cite{Kumar_2013}.\\
    \hline
    Inherited Weighted Methods Per Count & IWMC &  We proposed this notation for referring to inherited complexity, representing the sum of the complexities of a child class's parent classes. \\
    \hline
    Depth of Inheritance Tree & DIT & The Depth of Inheritance Tree is a software metric representing a class's depth in the inheritance hierarchy \cite{CK1994}. \\
    \hline
    Lack of Cohesion in Methods & LCOM & Lack of Cohesion in Methods denotes the degree of cohesion among the methods and attributes of a class. It is computed using the number of methods and the number of common instance variables \cite{CK1994}.\\
    \hline
    Hybrid Cyclomatic Complexity & HCC & The Hybrid Cyclomatic Complexity was introduced in \cite{Cernau2022}, and it is calculated by adding the WMC of the child class to the summed complexities of all the parent classes. \\
    \hline
  \end{tabular}
  \caption{Software Metrics Glossary}
\end{table}

\subsection{Motivation for selected algorithm and software metrics}
\label{subsection:motivation}

Throughout the whole software development lifecycle, the internal structure of the software system must be continuously assessed. Thus, software metrics are appropriate in this context for quantifying crucial components of the assessment. Software metrics can be used to automate the evaluation process and to detect software faults early in the system development process. 

The time and costs it takes to remedy software flaws incurred during system development are reduced when defects are predicted as early as possible\cite{hryszko2018}. During the maintenance and evolution of software systems, the issue of fault prediction is crucial. Software developers must constantly spot problematic software modules to raise the system's overall performance. However, machine learning-based classification models are currently being created to tackle the issue of defect prediction because it can be challenging to determine the circumstances under which a software module would have flaws.

Based on software quality measures, various machine learning methods, such as Decision Tree, Random Forest, Naive Bayes, or Fuzzy Inference Systems, can be used to find possibly flawed source code. The main advantage of employing these algorithms is that they automate the process of finding bugs and alerting programmers to possibly incorrect code, eliminating the need for humans to review metrics and form opinions based on them. Therefore, our machine learning-driven approach consists of extracting some features from the source codes and discriminating, based on them, whether the code contains at least one defect. The features are represented by a set of code metrics (WMC, LCOM, DIT, IWMC and HCC), while the discriminative algorithm is based on Support Vector Machines \cite{Vapnik}. The automatic defect identification performance is evaluated using the classical supervised learning quality criteria: accuracy, precision, and recall.

The primary motivation for choosing this set of metrics is to study the impact of the complexity of the code on the probability of the occurrence of defects. Therefore, we chose the WMC metric because it gives the sum of Cyclomatic Complexities for each class method \cite{Kumar_2013}, thus the overall complexity of a class. Moreover, to capture another aspect of complexity, we chose LCOM, a metric that defines the degree of cohesion of the elements of a class \cite{CK1994}. Since we wanted to study the importance of the inherited complexity of a class, the DIT metric \cite{CK1994} was the next choice. Based on this, we used the HCC, which defines the complexity of a class as the sum of its parents' complexities and its complexity. Finally, to have a better understanding of the inherited complexity, we defined the inherited complexity as inherited WMC (IWMC).
Support Vector Machine was the machine learning algorithm used to build the prediction model for these experiments. We chose this algorithm because the dataset and the feature set are small.

We employed three evaluation criteria for checking the correctness and performance of the categorisation process: precision, recall and accuracy. The classification model's precision represents the model's ability to identify the relevant items from the dataset (the number of classes that are accurately labelled as faulty divided by the total number of classes marked as defective). The classifier's recall or sensitivity aspect reveals the classifier's accuracy in detecting the faulty classes. Finally, the model's accuracy represents the ratio between the correct labelled items and the total number of predictions.

\begin{figure}[ht]
    \includegraphics[width=\columnwidth]{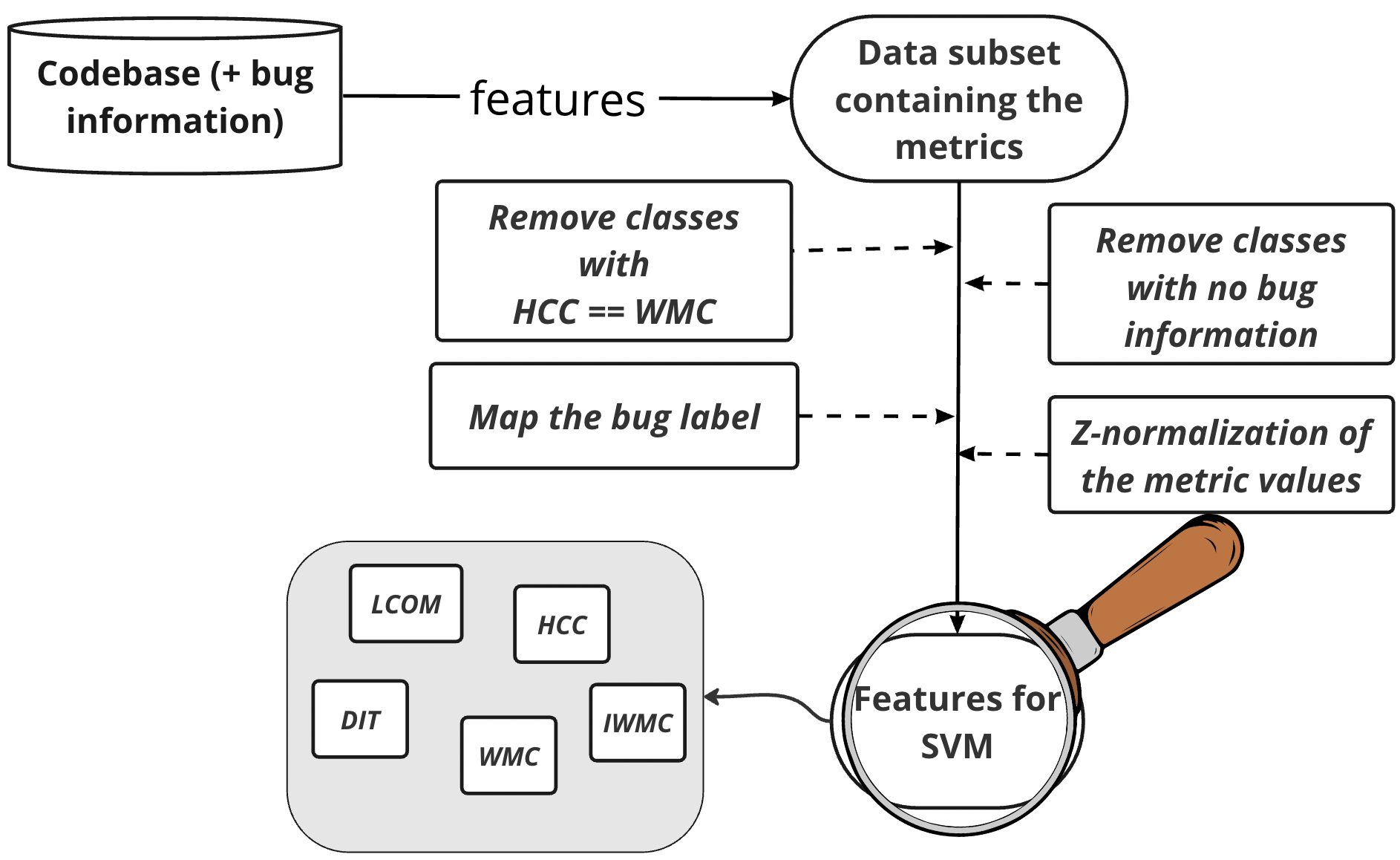}
    \caption[width=\columnwidth]{Data pre-processing flow}
    \label{fig:flow}
\end{figure}

Figure \ref{fig:flow} illustrates the process of transforming the raw data into relevant input data (features) for the Support Vector Machine algorithm to be trained and used as a defect prediction model. 
The steps performed on these datasets in order to be suitable inputs for the SMV algorithm are:
\begin{itemize}
    \item extracting and computing the values of the metrics that interest us (WMC, IWMC, HCC, LCOM, DIT) from the datasets.
    \item the classes with the HCC equal to WMC will be removed because we focus on studying the classes that inherit complexity.
    \item the classes with no bug information will be removed (\textit{no bug information} means there are no reported bugs linked to the respective file, meaning we do not know if they are correct or defective).
    \item the bug label will be binarised 
    , meaning all the values greater than one will be mapped to 1. Therefore, the bug label will only have two values, 0 or 1, meaning non-faulty and faulty.
\end{itemize}

As shown in figure \ref{fig:flow}, the focus will be on analysing the impact of the feature used for the prediction model and the relationships between them. More precisely, our approach aims to investigate two possible code source code representations: one that contains HCC, LCOM and DIT metrics ($R_1$) and one that contains WMC, IWMC, LCOM and DIT metrics ($R_2$) -- taking into account that HCC can be decomposed into WMC and IWMC.
 
In this research, we analyse the impact of inherited software complexity in the context of object-oriented programming since the early detection of defects in a software system can considerably affect the final stages of the development process. Specifically, we are examining the HCC feature's efficacy for automatic defect prediction compared to features that consider both inherited complexity and current class complexity as separate features. One of our aims is to determine whether the HCC measure is comprehensive and complex and does not minimise inheritance components. More precisely, to confirm its viability as a component of a prediction model. Another goal is to analyse how the child and inherited complexity perform as two separate features for defect prediction.

\subsection{Dataset description}
\label{subsection:dataset}

The dataset we used for the first experiments is a subset of a unified bug database \cite{Ferenc2018}, and it contained \textit{1151} Java files, from which \textit{211} files had bugs reported to them, and \textit{940} did not have any bugs reported to them. For simplicity, we will refer to this subset as the TinyBug dataset.
The second dataset we used for this paper is a Promise dataset \cite{PromiseDataset} for software defect prediction. In total, we used \textit{3473} Java files, from which \textit{1440} files had bugs reported to them, and \textit{2033} files did not have any reported defects related to them.
Finally, the third dataset used in this paper is the merged dataset, which is the union of these two previous datasets, and we will refer to it as the Unified dataset.

Regarding the TinyBug dataset, the authors of the dataset labelled classes with defects using different processes of mapping the defects reported in specific platforms with the existing classes in the dataset. On the other hand, we have no information about the labelling process for the Promise dataset. A summarisation of the datasets used for the experiments presented in this paper can be seen in Table \ref{tab:dataset-stats}.

\renewcommand{\arraystretch}{1.5}
\setlength{\arrayrulewidth}{1.2pt}

\begin{table}[h]
\centering
\begin{tabular}{|l|c|c|c|}
\hline
\textbf{Dataset} & \textbf{Total Files} & \textbf{Faulty files}& {\textbf{Non-faulty files*}} \\
\hline
TinyBug Dataset & 1151 & 211 & 940 \\
\hline
Promise Dataset & 3473 & 1440 & 2033 \\
\hline
Unified Dataset & 4624 & 1651 & 2973 \\
\hline 
\multicolumn{4}{l}{\textsuperscript{*} Non-faulty files have no reported bugs related to them.}
\end{tabular}
\caption{Dataset Statistics}
\label{tab:dataset-stats}
\end{table}

\subsection{Exploratory data analysis}
\label{subsection:EDA}

This section focuses on analysing the relationships and correlations between the features of the prediction model. Therefore, the first data analysis focuses on the distribution of the files (faulty vs non-faulty). This was done using a KDE (Kernel Density Estimate) plot and by calculating the mean (average) number of bugs and the standard deviation. The following analysis observes the correlations between the considered features (the software metrics, HCC, LCOM, DIT, IWMC, and WMC). 

We used the WMC, DIT, and LCOM metrics as features for the SVM algorithm. In addition, we computed the HCC values for each file and added a new column, IWMC, which is the inherited WMC for each file. Each dataset file has a column for bug information attached to it. 

The analysis presented here is conducted using the preprocessed initial dataset (the classes with the WMC equal to the HCC were removed from the dataset). Additionally, all analyses are performed after normalising the bug number values in each file, converting them into binary categories of 1 or 0.

One thing worth mentioning here is that this approach is agnostic regarding the type of architecture of the project (mobile/desktop/web) of which the studied classes are part. Since these metrics are calculated at the class level, the study is not affected by the specific characteristics of each type of project.

\subsubsection{Distribution of faulty and non-faulty files}
\label{subsubsec:bug-distribution}
\begin{figure}
\centering
\begin{minipage}{.5\textwidth}
  \centering
  \includegraphics[scale=0.45]{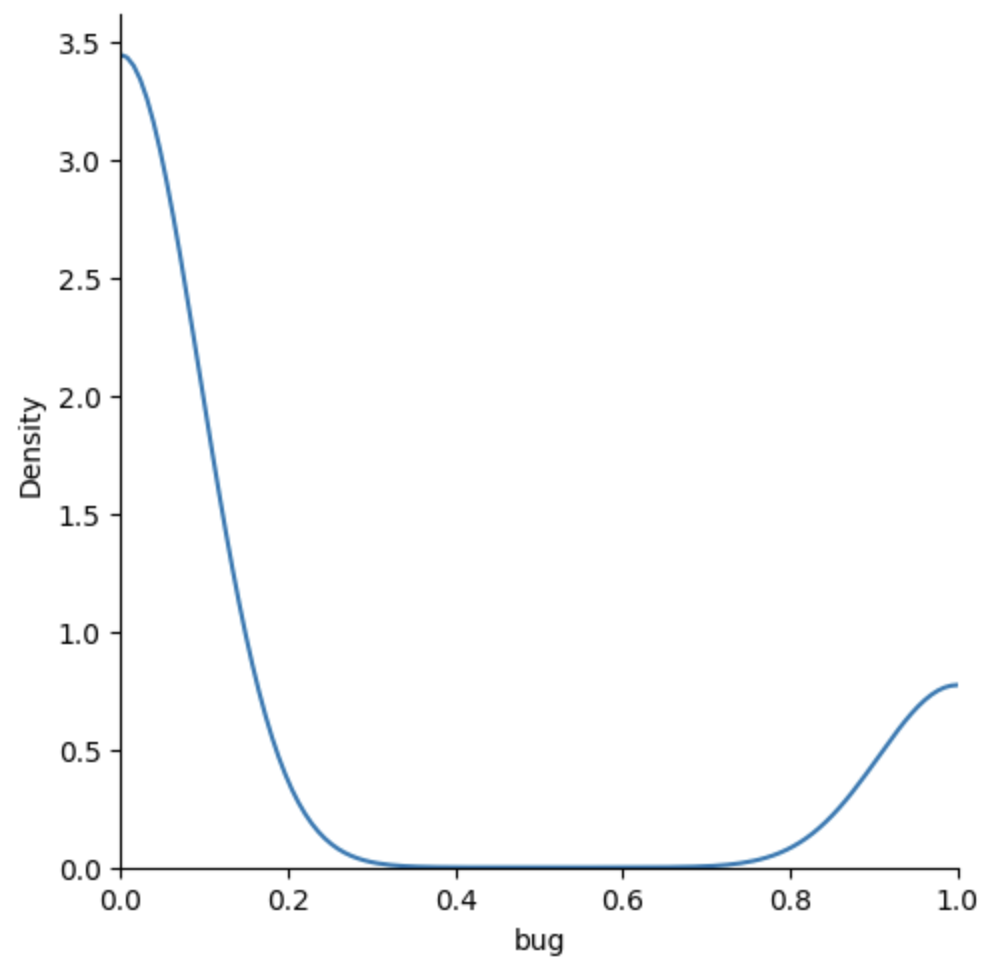}
  \captionof{figure}{Defect density in the \\ TinyBug dataset}
  \label{fig:defect-density-tiny}
\end{minipage}%
\begin{minipage}{.5\textwidth}
  \centering
  \includegraphics[scale=0.45]{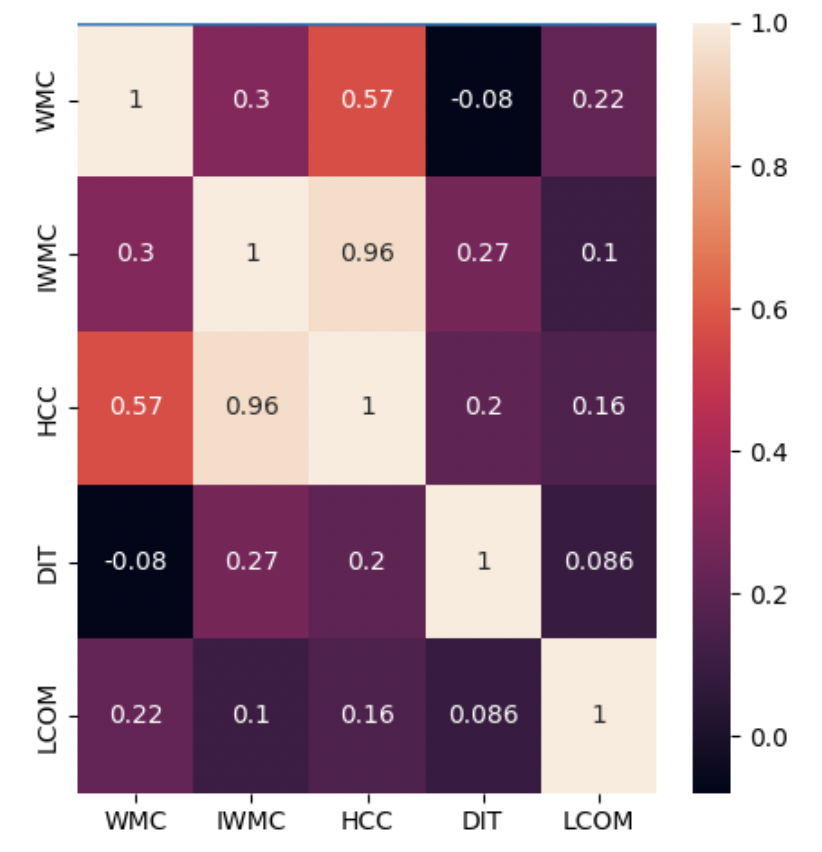}
  \captionof{figure}{Software metrics correlations \\ in the TinyBug dataset}
  \label{fig:metrics-correlation-tiny}
\end{minipage}
\end{figure}

\begin{figure}
\centering
\begin{minipage}{.5\textwidth}
  \centering
  \includegraphics[scale=0.45]{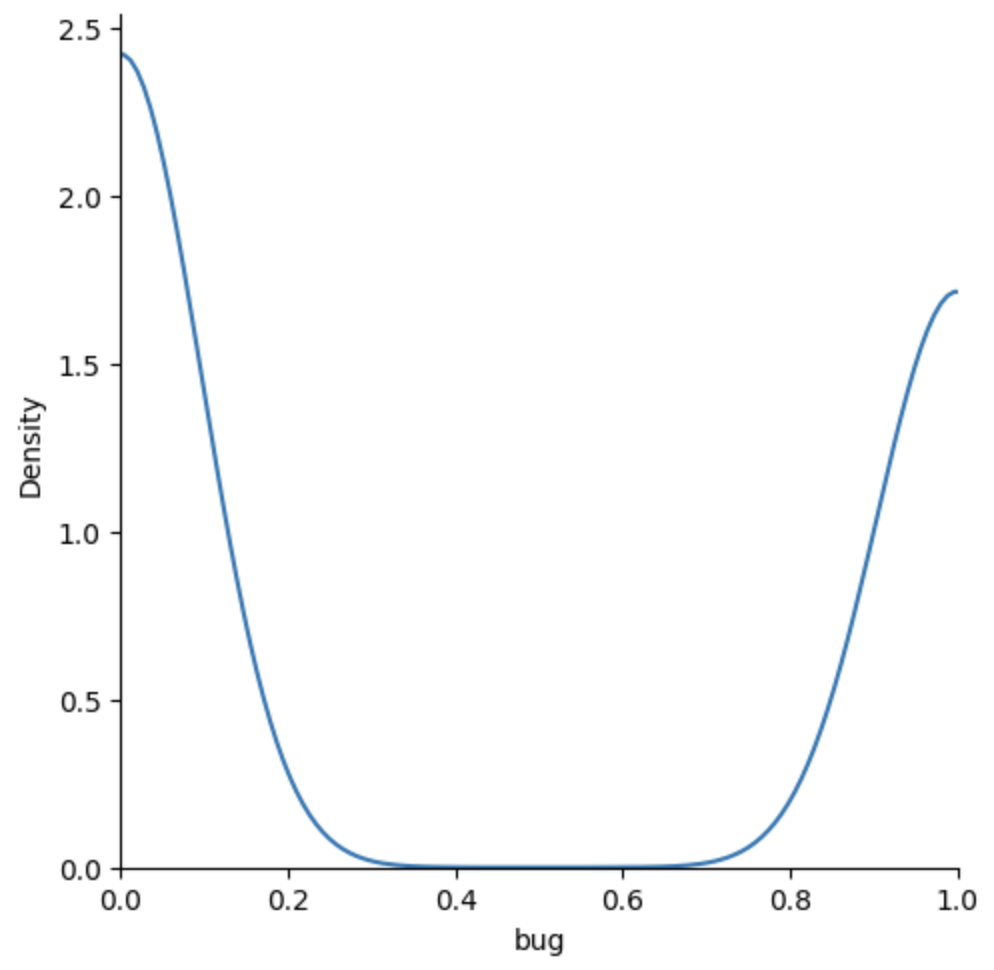}
  \captionof{figure}{Defect density in the \\ Promise dataset}
  \label{fig:defect-density-promise}
\end{minipage}%
\begin{minipage}{.5\textwidth}
  \centering
  \includegraphics[scale=0.45]{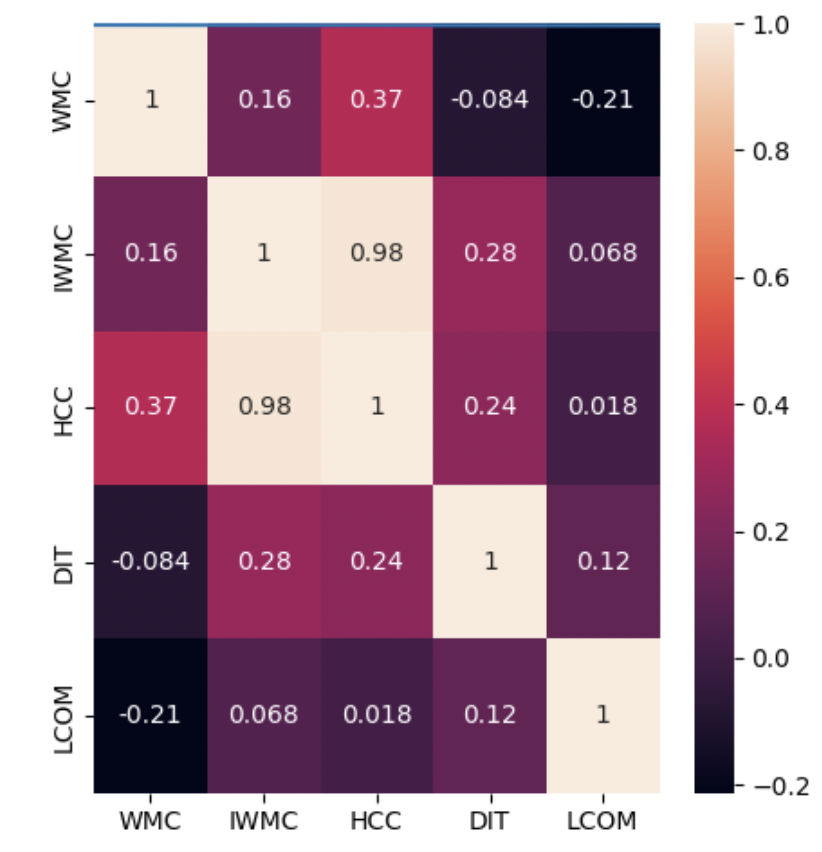}
  \captionof{figure}{Software metrics correlations \\ in the Promise dataset}
  \label{fig:metrics-correlation-promise}
\end{minipage}
\end{figure}

\begin{figure}
\centering
\begin{minipage}{.5\textwidth}
  \centering
  \includegraphics[scale=0.45]{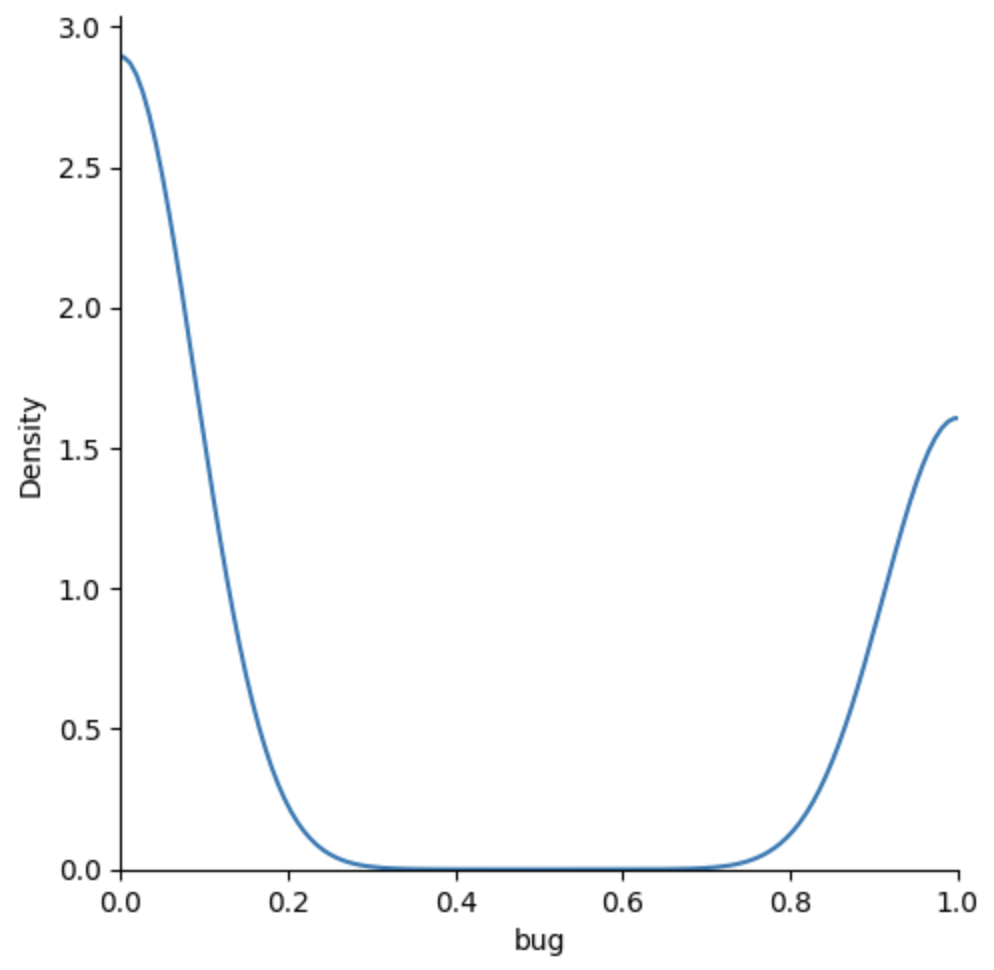}
  \captionof{figure}{Defect density in the \\ Unified dataset}
  \label{fig:defect-density-merged}
\end{minipage}%
\begin{minipage}{.5\textwidth}
  \centering
  \includegraphics[scale=0.45]{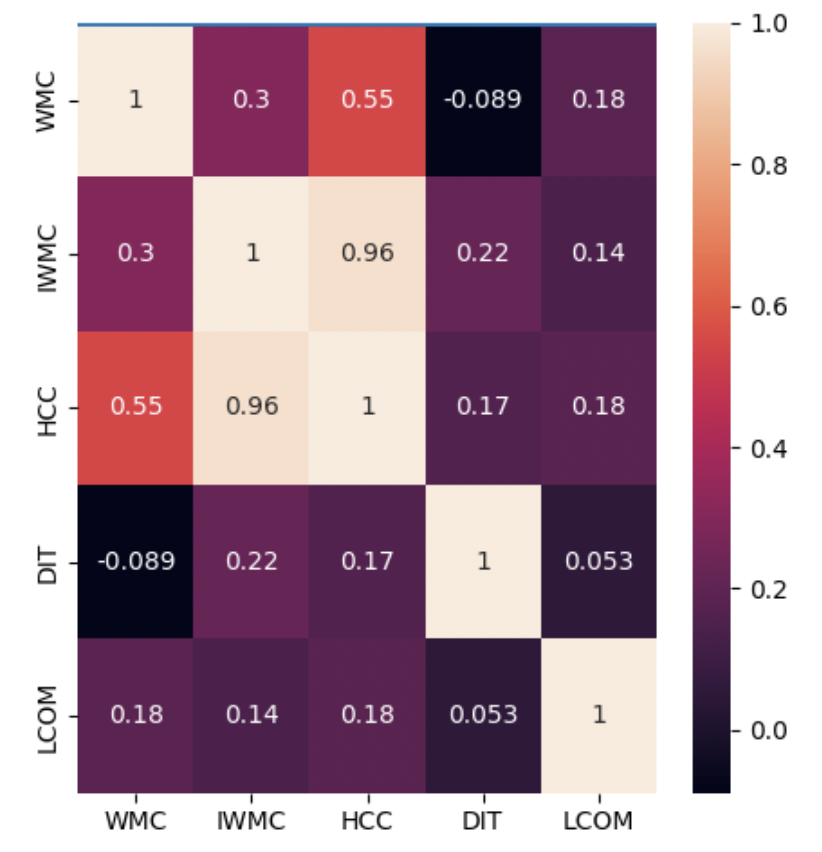}
  \captionof{figure}{Software metrics correlations \\ in the Unified dataset}
  \label{fig:metrics-correlation-merged}
\end{minipage}
\end{figure}

In order to analyse our data, we visualise the distribution of the bugs over the source code files. In figures \ref{fig:defect-density-tiny}, \ref{fig:defect-density-promise}, and \ref{fig:defect-density-merged}, the density of defects in the three datasets can be seen. The defect density was computed using the Seaborn Python library, more precisely, the displot function \cite{SeabornDistplot}. 
The function uses Kernel density estimation (KDE) to estimate the probability density function of a random variable.

The figures (\ref{fig:defect-density-tiny}, \ref{fig:defect-density-promise} and \ref{fig:defect-density-merged}) show the density of the faulty and non-faulty files in each of the three datasets. In each case, the peak for the non-faulty files is higher than for the faulty ones, meaning that all the datasets have a higher number of files with no reported bugs. However, the Promise and the Unified datasets exhibit a relatively balanced distribution of the bug data, whereas for the Tiny dataset, the difference is significant. The proportions of the faulty and non-faulty files can be seen in table \ref{table:avg-std-bugs}.

\def\arraystretch{1.5}
\setlength\tabcolsep{5pt}
\begin{table}[h!]
\centering
\begin{tabular}[t]{|c|c|c|}
\hline
\textbf{Dataset}&\textbf{Non-faulty files}&\textbf{Faulty files}\\
\hline
TinyBug dataset&81,67\%&18.33\%\\
\hline
Promise dataset&58,54\%&41.46\%\\
\hline
Unified dataset&64,29\%&35.71\%\\
\hline
\end{tabular}
\caption{Proportion of non-faulty and faulty files}
\label{table:avg-std-bugs}
\end{table}

\subsubsection{Features correlations}
\label{subsubsec:correlations}
Next, we highlight the possible correlations among the considered features (in Figures \ref{fig:metrics-correlation-tiny}, \ref{fig:metrics-correlation-promise} and \ref{fig:metrics-correlation-merged}) since we are interested in identifying the feature importance in the automatic defect identification process.

Thus, the heatmap from Figure \ref{fig:metrics-correlation-promise} indicates the correlation among the considered software metrics (WMC, IWMC, HCC, LCOM, DIT) in the Promise dataset through the Pearson's R values (that measure how each feature is correlated to one another: -1 signifies zero correlation, while 1 signifies a perfect correlation). We can notice that the features involved in the first representation (LCOM, DIT, HCC) are not correlated, which is essential for the performance of the classification model. Furthermore, the integrity of the classifier is not compromised in the second case since the features involved in the second representation (LCOM, DIT, IWMC, WMC) are independent.  Similar trends are identified for the other two datasets, as seen in \ref{fig:metrics-correlation-tiny} and \ref{fig:metrics-correlation-merged}. When we analyse the correlation among HCC, WMC and IWMC, we notice that HCC and IWMC are strongly correlated. Still, the correlation between HCC and WMC is not very strong (0.33), which means that even if the HCC value is computed using WMC, there is no strong correlation between them. More precisely, HCC can be used as an independent feature in a prediction model.

Furthermore, the IWMC and WMC exhibit a degree of independence. Also, IWMC, WMC, DIT, and LCOM are independent, supporting our first research question about investigating the power of the two representations $ R_1 $ and $ R_2 $.

\section{Experiments}
\label{section:experiments}
Based on the previous analysis (described in Section \ref{subsection:EDA}), we aim to explore the capacity of a machine learning algorithm to identify the source code defects automatically. The algorithm will consider the computed software metrics as features for all three mentioned datasets.

We trained and validated a classification model based on each data representation for all three datasets. The first data representation includes the HCC, LCOM, and DIT metrics. This scenario performed the best in some preliminary experiments \cite{Cernau2022}. The second set of features was the decomposition of the HCC metric into IWMC and WMC, together with LCOM and DIT. 

The values for the features have been normalised by performing a Z-normalization \cite{mitchell1997mcgraw} using \textit{StandardScaler} from the \textit{sklearn} Python library. The algorithm used for our binary classification problem (with or without bug) was \textit{C-Support Vector Classification} \cite{SVC} algorithm from the \textit{SVM sklearn module}. The algorithm was run with the linear kernel type, and the value for the \textit{C} parameter was the default one, 1.0. 

The data distribution for test and train subsets was made using the sampling method from Pandas DataFrame in Python with frac=0.7 and random state=1. The input data was split into train and test data using the 70/30 rule (70\% of data is used for training the prediction model, and 30\% of data is used for testing the model). The train and test data sets have approximately 50\% files with bugs and 50\% without bugs, randomly chosen.  

\subsection{Model prediction performance evaluation}
The evaluation of these experiments for defect prediction can be seen in tables \ref{table:scores-HCC-LCOM-DIT} and \ref{table:scores-WMC-LCOM-DIT-IWMC} for each of the three datasets. Both representations have been used in these experiments. The criteria by which the correctness and performance of the prediction model were evaluated are precision, recall and accuracy.

\def\arraystretch{1.5}
\begin{table}[htbp]
\centering
\tiny
\begin{tabular}{|c|c|c|c|c|c|c|c|}
\hline
\multirow{2}{*}{Evaluation criteria} & \multicolumn{2}{c|}{Tiny dataset} & \multicolumn{2}{c|}{Promise dataset} & \multicolumn{2}{c|}{Unified dataset} \\ \cline{2-7} 
& Faulty & Non-Faulty & Faulty & Non-Faulty & Faulty & Non-Faulty \\ \hline
Precision & 0.6666 & 0.5277 & 0.68 & 0.5153 & 0.6934 & 0.5310 \\ \hline
Recall & 0.1904 & 0.9047 & 0.1066 & 0.9498 & 0.1919 & 0.9151 \\ \hline
Accuracy & \multicolumn{2}{c|}{0.5476} & \multicolumn{2}{c|}{0.5282} & \multicolumn{2}{c|}{0.5535} \\ \hline
\end{tabular}
\caption{Classification Scores for the HCC-LCOM-DIT representation}
\label{table:scores-HCC-LCOM-DIT}
\end{table}

\begin{table}[htbp]
\centering
\tiny
\begin{tabular}{|c|c|c|c|c|c|c|c|c|}
\hline
\multirow{2}{*}{Evaluation criteria} & \multicolumn{2}{c|}{Tiny dataset} & \multicolumn{2}{c|}{Promise dataset} & \multicolumn{2}{c|}{Unified dataset} \\ \cline{2-7} 
& Faulty & Non-Faulty & Faulty & Non-Faulty & Faulty & Non-Faulty \\ \hline
Precision & 0.75 & 0.5471 & 0.6314 & 0.5319 & 0.6434 & 0.5434 \\ \hline
Recall & 0.2380 & 0.9206 & 0.2471 & 0.8557 & 0.2989 & 0.8343 \\ \hline
Accuracy & \multicolumn{2}{c|}{0.5793} & \multicolumn{2}{c|}{0.5514} & \multicolumn{2}{c|}{0.5666} \\ \hline
\end{tabular}
\caption{Classification Scores for WMC-LCOM-DIT-IWMC representation}
\label{table:scores-WMC-LCOM-DIT-IWMC}
\end{table}

The \textit{WMC-LCOM-DIT-IWMC} scenario had increased accuracy for all three datasets. Moreover, the recall for the faulty entries for all the datasets in the \textit{WMC-LCOM-DIT-IWMC} scenario is considerably improving, meaning that the ability of the model to detect the positive items better is improving. Considering these, we can answer the first research question, \textit{How do the two features (the class's inherited and actual complexity) impact the automatic identification of software defects?}. As was shown in the previous sections, the HCC and IWMC metrics are not correlated. Therefore, they can be used as features in a classification algorithm. Moreover, we can see better results when they are used as different features rather than when only the HCC metric is used.

\subsection{Comparison between the \textit{WMC-LCOM-DIT-IWMC} and \textit{HCC-LCOM-DIT} scenarios on concrete example}

An example of a class for which the prediction of the probability of occurrence of bugs was different is class \textit{org.apache.tools.ant.taskdefs.Replace}, from the PROMISE dataset, the \textbf{ant} library, version 1.5.
The class has the following values for the features:
\begin{itemize}
    \item HCC = 31
    \item LCOM = 0.7777
    \item DIT = 4
    \item IWMC = 21
    \item WMC = 19
    \item bug = 1 
\end{itemize}

For these features, the \textit{WMC-LCOM-DIT-IWMC} and \textit{HCC-LCOM-DIT} scenarios have different predictions for the bug probability. More precisely, the \textit{WMC-LCOM-DIT-IWMC} scenario correctly predicted that the class contains a bug. At the same time, the \textit{HCC-LCOM-DIT} scenario did not predict the bug's presence.

This example supports the hypothesis that inherited complexity can have an impact on the erroneous quality of a class. To answer the second research question (\textit{What inheritance-related aspects does the Hybrid Cyclomatic Complexity measure try to emphasise?}), we can say that the HCC metric tries to capture the impact of inherited complexity on the child class. However, as we can see, the \textit{WMC-LCOM-DIT-IWMC} performs better than the \textit{HCC-LCOM-DIT} scenario, more precisely, when the inherited complexity is used as a stand-alone feature.

\section{Threats to validity}
\label{section:threats-to-validity}

\subsection{Internal validity}
We assumed that if a class is part of a deep inheritance tree, its error-prone probability is higher due to all the inherited complexity. For example, code listings \ref{lst:address-transformer}, \ref{lst:customer-transformer}, \ref{lst:order-transformer} and \ref{lst:order-details-transformer} represent a suite of four entity transformers, where the last child, \textit{OrderDetailsTransformer}, contains a logic bug. The values for the HCC, LCOM, DIT, IWMC, and WMC metrics computed using the MetricsReloaded Intellij plugin are shown in table \ref{table:threats-to-validity}. Moreover, we tested the \textit{OrderDetailsTransformer} against the prediction model, and the prediction was that it did not contain a defect. 

\def\arraystretch{1.5}
\setlength\tabcolsep{5pt}
\begin{table}[h!]
\centering
\begin{tabular}[t]{cccccc}
\hline
Class&HCC&LCOM&DIT&IWMC&WMC\\
\hline
OrderDetailsTransformer&4&1&4&3&1\\
\hline
OrderTransformer&3&1&3&2&1\\
\hline
CustomerTransformer&2&1&2&1&1\\
\hline
AddressTransformer&1&1&1&0&1\\
\hline
\end{tabular}
\caption{MetricsReloaded results for OrderDetailsTranformer, OrderTranformer, CustomerTransformer and AddressTransformer}
\end{table}
\label{table:threats-to-validity}

\begin{lstlisting}[caption=AddressTransformer, label={lst:address-transformer}]
public class AddressTransformer {
    public String transformAddress(Address address) {
        return "The address will be delivered to the following address: "
                + address.getNumber() + " " + address.getStreet() +", " + address.getCity() + ", "
                + address.getCountry() + "\n";
    }
}
\end{lstlisting}

\begin{lstlisting}[caption=CustomerTransformer, label={lst:customer-transformer}]
public class CustomerTransformer extends AddressTransformer {
    public String transformCustomer(Customer customer) {
        return "The contact person for this order is "
                + customer.getName() + " " + customer.getSurname() + "\n"
                + this.transformAddress(customer.getAddress());
    }
}
\end{lstlisting}

\begin{lstlisting}[caption=OrderTransformer, label={lst:order-transformer}]
public class OrderTransformer extends CustomerTransformer{
    public String transformOrder(Order order){
        return "Order reference number " + order.getId() + "\n" +       this.transformCustomer(order.getCustomer());
    }
}
\end{lstlisting}

\begin{lstlisting}[caption=OrderDetailsTransformer, label={lst:order-details-transformer}]
public class OrderDetailsTransformer extends OrderTransformer{
    public String getOrderSummary(Order order) {
        return this.transformOrder(order) + this.transformCustomer(order.getCustomer());
    }
}
\end{lstlisting}

\subsection{External validity}
The main external validity thread is caused by the fact that in the experiments performed in this paper, we used datasets from two different sources. More precisely, the main concern is how the metrics were calculated and whether the exact definition was used for the two datasets. On the one hand, we know that for the \textit{TinyDataset}, the authors used Checkstyle \cite{Checkstyle} to calculate the values of the metrics. However, we did not find relevant information to specify how the metrics for the \textit{Promise dataset} were calculated. Therefore, we are not sure that the same metric definitions were used.

\section{Future work}
\label{section:future_work}

In terms of future work, we want to analyse further the consequences of the inherited complexity on the complexity of a class. However, the experiments done in this paper show that the inherited complexity is not correlated with the actual complexity of the class. Therefore, they both can be used as features in a prediction model. Moreover, we should analyse the efficiency of this approach for predicting other software quality attributes. In addition, by analysing its two component parts, we can better define the HCC metric and convert it into a robust and complex software metric.
Moreover, this study does not consider the composition relationship between the classes. Since the inheritance versus composition subject is well-known in the software engineering domain, one future step is to analyse if we can apply the same decomposition process for other software metrics (related to class composition). Another study subject can be the nested classes and how they impact the overall complexity.
Finally, a more comprehensive topic of analysis is the impact of improving complexity on other quality attributes, such as cohesion. The present study studies the implications of software complexity in isolation without considering its effects.

\section{Conclusion}
\label{section:concl}
Software complexity has an essential role in the software development lifecycle of a project, but mainly in the testing and maintaining phases. Therefore, measuring and managing the software complexity from the early stages of a project is mandatory. In addition, hot spots in a software system can be identified using software complexity metrics, with the highest probability of generating defects and additional costs in the future.

The HCC was introduced as a software metric that defines the complexity of a class as the sum of its complexity and its parents' complexities. The data analysis and experiments performed in this paper show that the inherited complexity is not correlated to the actual complexity of the class. This further proves that the HCC metric is a valid software metric. However, the evaluation of the effectiveness of the HCC versus the inherited complexity as a metric showed that they are both powerful features in a defect prediction model. This means that inherited complexity is indeed a power factor that needs to be considered when speaking about the complexity of a system.

\section{Declaration of generative AI and AI-assisted technologies in the writing process}
During the preparation of this work the authors used \textit{Grammarly} in order to check and correct the written text in terms of grammar and expression. After using this tool, the authors reviewed and edited the content as needed and take full responsibility for the content of the publication.

\bibliographystyle{plain}
\bibliography{BibAll}
\noindent
\begin{minipage}[t]{0.3\textwidth}
    \vspace{0pt}
    \includegraphics[width=\textwidth]{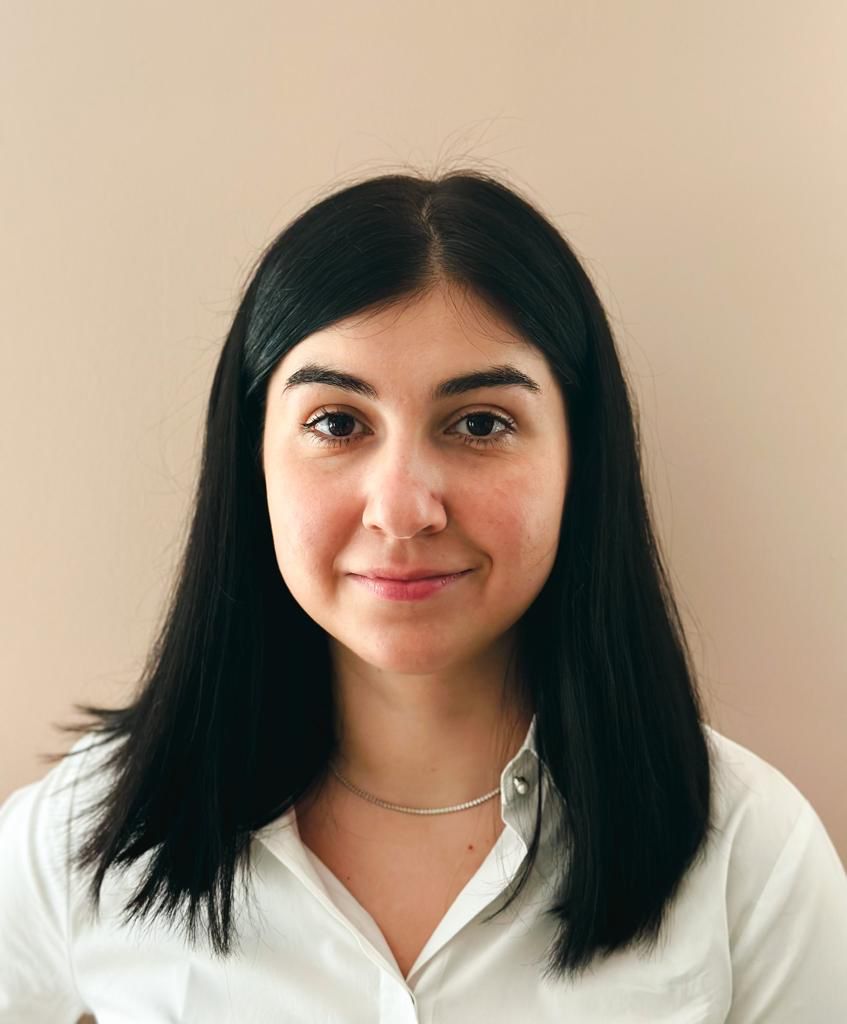}
    \centering
\end{minipage}
\hfill
\begin{minipage}[t]{0.65\textwidth}
    \vspace{0pt}
    \raggedright
Laura Diana Cernău is a PhD student at Babeș-Bolyai University, Faculty of Mathematics and Informatics, with over seven years of professional experience as a software engineer. Her expertise spans technical ownership, implementation, platform monitoring, and designing domain architectures and strategies.

Her PhD research focuses on analyzing the correlation between software metrics and critical software attributes such as maintainability, testability, and reliability. By connecting these metrics to the characteristics of well-structured software systems, her work aims to predict error-prone or defective design entities. To achieve this, she employs advanced artificial intelligence algorithms and methods to validate and demonstrate these relationships. Her research contributes to improving the quality and robustness of software systems.
\end{minipage}

\vspace{1cm} 

\noindent
\begin{minipage}[t]{0.3\textwidth}
    \vspace{0pt}
    \includegraphics[width=\textwidth]{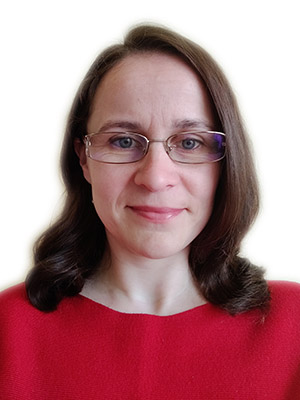}
    \centering
\end{minipage}
\hfill
\begin{minipage}[t]{0.65\textwidth}
    \vspace{0pt}
    \raggedright
Laura Dioșan is currently a full professor in the Department of Computer Science at UBB. She has a vast research experience participating in more projects in the last 15 years, from which she coordinated five projects successfully completed. Her research areas cover evolutionary optimisation, swarm intelligence, and machine learning, with excellent research results obtained by different hybrid models given by evolutionary computation and kernel algorithms for image processing and classification.

Another part of her research activity has been dedicated to self-adaptation of algorithms to the problem to be solved (in order to improve the quality of the solving/learning process) through parameter optimisation and the fusion of information. Research results have been published in journals and conference papers, obtaining more than 900 citations.
\end{minipage}

\vspace{1cm} 

\noindent
\begin{minipage}[t]{0.3\textwidth}
    \vspace{0pt}
    \includegraphics[width=\textwidth]{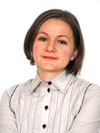}
    \centering
\end{minipage}
\hfill
\begin{minipage}[t]{0.65\textwidth}
    \vspace{0pt}
    \raggedright
Camelia Şerban is an Associate Professor in the Department of Computer Science at Babeș-Bolyai University, Romania. She is an enthusiastic and humorous teacher and facilitator with more than 20 years of experience in higher education.

Camelia delivers university courses on Advanced Programming Methods and Data Structures. Her research interests include Software Metrics, Software Assessment, and Object-Oriented Design. She is particularly interested in studying aspects related to software quality. Her potential is greatest when working in a team. See \texttt{http://www.cs.ubbcluj.ro/\~camelia} for details.
\end{minipage}

\end{document}